\documentclass[runningheads]{llncs}
\usepackage[T1]{fontenc}
\usepackage{graphicx}
\usepackage{cite}
\usepackage{float}

\begin{document}
\title{Automated Segmentation of Ischemic Stroke Lesions in Non-Contrast Computed Tomography Images for Enhanced Treatment and Prognosis}
\titlerunning{Automated Segmentation of Ischemic Stroke Lesions}
%
\author{Toufiq Musah\inst{1},
Prince Ebenezer Adjei \inst{1,2},
Kojo Obed Otoo \inst{3}}
\authorrunning{Toufiq Musah et al.}
%
\institute{Kwame Nkrumah University of Science and Technology, Department of Computer Engineering
\\\and
Kumasi Centre for Collaborative Research in Tropical Medicine\\\and
Komfo Anokye Teaching Hospital, Radiology Department}
\maketitle              
\begin{abstract}
Stroke is the second leading cause of death worldwide, and is increasingly prevalent in low- and middle-income countries (LMICs). Timely interventions can significantly influence stroke survivability and the quality of life after treatment. However, the standard and most widely available imaging method for confirming strokes and their sub-types, the NCCT, is more challenging and time-consuming to employ in cases of ischemic stroke. For this reason, we developed an automated method for ischemic stroke lesion segmentation in NCCTs using the nnU-Net frame-work, aimed at enhancing early treatment and improving the prognosis of ischemic stroke patients. We achieved Dice scores of 0.596 and Inter-section over Union (IoU) scores of 0.501 on the sampled dataset. After adjusting for outliers, these scores improved to 0.752 for the Dice score and 0.643 for the IoU. Proper delineation of the region of infarction can help clinicians better assess the potential impact of the infarction, and guide treatment procedures.

\end{abstract}

\section{Introduction}
There is an estimated 15 million cases of stroke annually, with one-third of these cases leading to death, and another one-third leading to some form of disability \cite{feigin2017global}. The rate of prevalence is more apparent in Low- and Middle-Income Countries (LMICs), where existing healthcare infrastructures remain unprepared to deal with the surge in cases \cite{Prevalence-Stroke, Epidemic-Stroke}. Timely intervention is a critical factor in the survivability and post-treatment quality of life of patients \cite{Time-Critical}. Expert radiologist interpretations of the standard Non-Contrast Computed Tomography (NCCT) images are far more accurate for hemorrhagic stroke (95\%), when compared to ischemic stroke where a diagnostic ruling is given for major onsets only two-thirds of the time \cite{CT-Accuracy, Time-Critical}. CT imaging is the most accessible modality for brain diagnostic imaging in LMICs \cite{Image-Inventory}.

The process of diagnosing stroke, and subsequently segmenting the lesion region using radiological images can guide treatment and predict rehabilitation outcomes \cite{Standard-Manual-Segment, Lesion-Biomarkers}. Manual segmentation can be highly time-consuming and error-prone, with significant inter-observer variability depending on the observer's level of expertise \cite{Manual-Seg-Errors}. The process of segmenting lesions in NCCT images is a relatively difficult task (due to low tissue contrast) as compared to MRI \cite{Difficult-CT}, especially when done manually. It is important that robust automated lesion segmentation processes for varying modalities (in this case, NCCT) be made available, especially in regions with limited imaging modality accessibility.

Several works explore the segmentation of stroke lesions, employing a variety of approaches. Liang et. al \cite{Liang_2021} proposed the Symmetry-Enhanced Attention Network (SEAN) to automatically segment acute ischemic infarcts in NCCTs. The input image is first transformed into the standard space in an unsupervised manner by an 'Alignment Network'. In the transformed image with its bilateral symmetry, long range dependencies are better captured by SEAN, which has proven to outperform other existing symmetry based methods, with a dice score of 0.5784. The nnU-Net was developed by Fabian et. al \cite{nnUNet}, and is described as an out-of-the-box tool for biomedical image segmentation. It is a well-validated segmentation method, and has achieved state-of-the-art (SOTA) results in various biomedical image segmentation benchmark datasets such as the Medical Segmentation Decathlon (MSD) \cite{MSD-Data}, Brain Tumor Segmentation (BraTS) Challenge \cite{Brats-Challenge}, and the Ischemic Stroke Lesion Segmentation Challenge (ISLES) \cite{ISLES-Challenge}.

In this paper, we achieve promising results in ischemic lesion segmentation in NCCTs, by using out-of-the-box, well-validated methods. Initial image processing involved normalizing the input NCCT scans to ensure uniformity in image quality and resolution. We employed a Residual Encoder U-Net architecture from the nnU-Net framework, known for its robust performance in medical image segmentation tasks. The network was trained using a curated dataset consisting of NCCT images which were annotated using DWI MRI scans of the subjects as reference. The results obtained reiterate the potential of using validated methods to automate the difficult task of ischemic lesion segmentation in NCCT scans.  Comparisons with existing techniques highlight the improvements and underscore areas for future enhancements.

\section{Materials and Methods}
\subsection{Dataset}
\begin{figure}
\includegraphics[width=\textwidth]{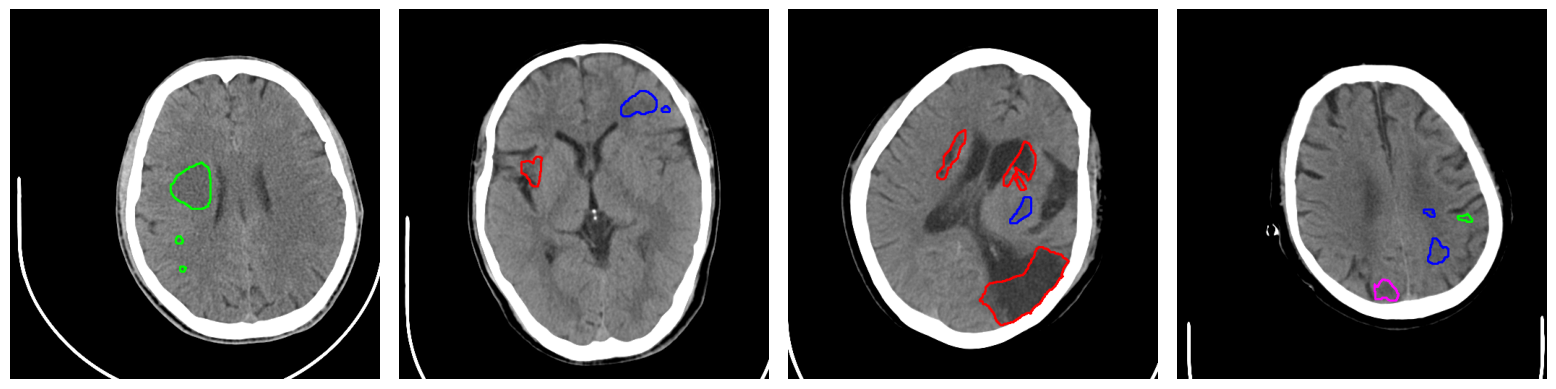}
\caption{Segmentation masks indicating different types of infarcts in the scans. Each color represents a specific type of infarct: Red - Clear acute infarcts, Blue - Remote infarcts, Green - Blurred acute infarcts, Magenta - Infarct.} \label{fig:infarcts}
\end{figure}

The dataset used for this study is the Acute Ischemic stroke Dataset (AISD) \cite{Liang_2021}, comprising of Non-Contrast-enhanced Computed Tomography (NCCT), and diffusion-weighted MRI (DWI) scans from 398 subjects. The NCCT scans are obtained less than 24 hours from the onset of ischemia symptoms, and have a slice thickness of 5mm. The lesion regions in the NCCT are manually segmented by a doctor using the DWI scans (taken within 24 hours after the NCCT imaging procedure) as reference, and then double-reviewed by a senior doctor. The scans are multi-labelled for various forms of ischemias present as shown in Figure \ref{fig:infarcts}. This study only considers acute infarctions, as they are more pertinent in onset strokes, and require immediate response in diagnosis and treatment.

\subsection{Data Preprocessing}
The data preprocessing pipeline was automatically selected and applied by nnU-Net, using its new ResEnc presets \cite{isensee2024nnu}. The first step involved CT Normalization, which clips the image intensities between the 0.5th and 99.5th percentiles to remove outlier pixels, followed by zero-mean and unit-variance normalization based on the intensity statistics of the dataset. It standardizes the images which may have been produced using varying protocols and or machinery, while preserving relevant CT characteristics. A small epsilon value of $1 \times 10^{-8}$ was used to prevent division by zero when standardizing.

The individual 5mm slices were handled as 2D images and resampled  to an isotropic spacing of 1mm x 1mm, with patch sizes of 512 x 512 pixels. For image data, cubic interpolation (order 3) was used for in-plane resampling, while linear interpolation (order 1) was applied to segmentation masks to preserve label integrity.

\subsection{Model Architecture}
In using the out-of-the-box nnU-Net framework for both preprocessing and training, the selected architecture was the Residual Encoder U-Net, which was first proposed by \cite{isensee2019attempt}, in a 3D configuration. It comprised of 8 stages in the encoder (with the following feature maps per stage: [32, 64, 128, 256, 512, 512, 512, 512]), and 7 stages in the decoder. In the encoder, the first three stages had 1, 3, and 4 blocks respectively, whiles the last five stages each had 6 blocks. The decoder only had 1 block per stage. This design allows for increasing feature complexity as the network deepens.

\begin{figure}
\includegraphics[width=\textwidth]{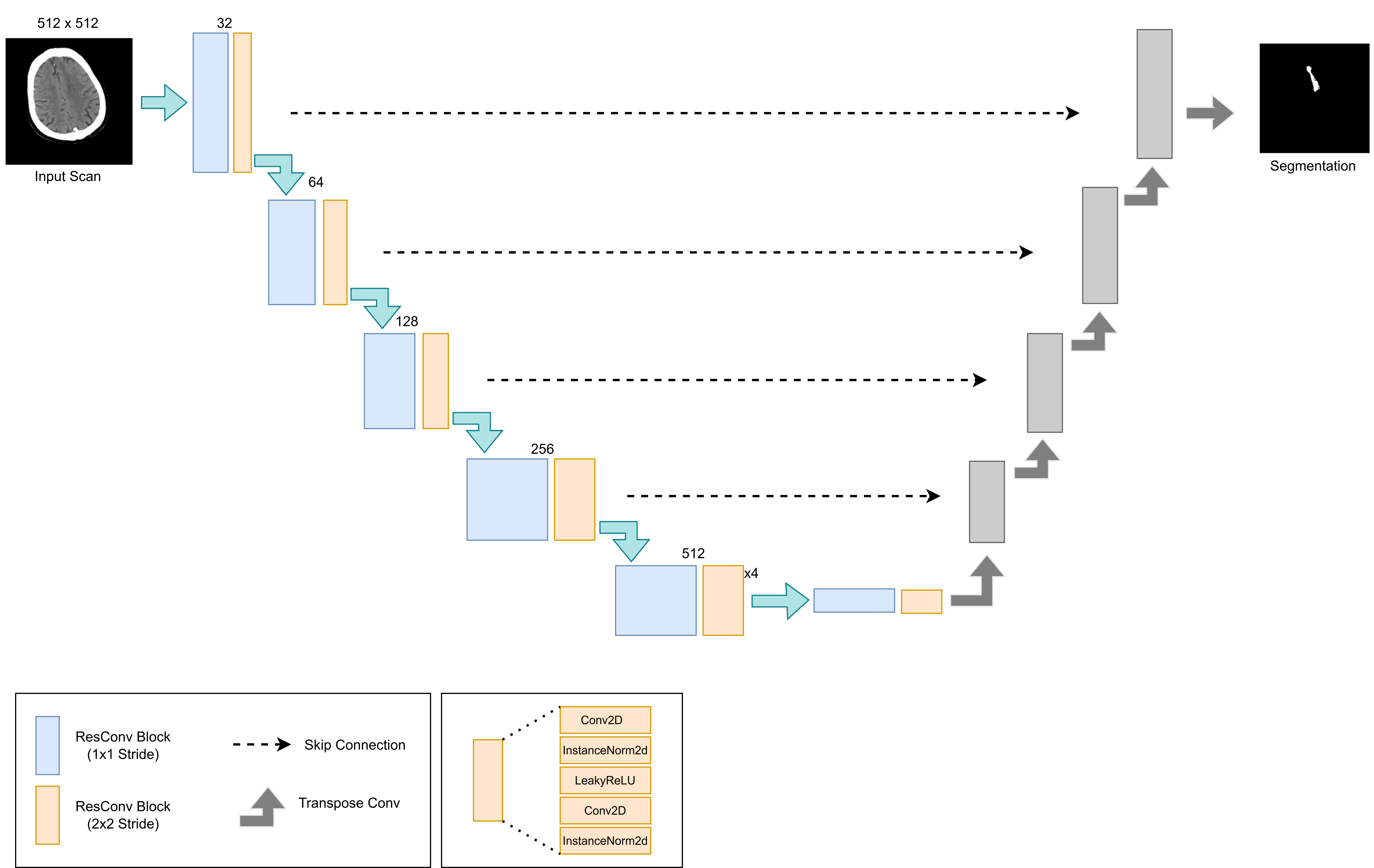}
\caption{Residual Encoder U-Net Architecture} \label{fig:resenc}
\end{figure}

Each residual convolution block in the encoder comprised of 5 layers as seen in Figure \ref{fig:resenc}. An input 2D convolution layer, followed by an instance normalization, a LeakyReLU activation function, another 2D convolution layer, and finally an instance normalization of the input features before handing it off to the next block.

\subsection{Training Recipe}
The network was trained for 50 epochs using a linearly decaying learning rate starting from 0.01. We used a batch size of 13 and calculated dice losses per batch. Five-fold cross-validation was performed, with testing conducted on two fold sets. The entire training process took approximately 4 hours using an NVIDIA T4 GPU.

\section{Results and Discussion}
Acute infarct lesions in the non-contrast computed tomography (NCCT) scans were segmented utilizing the trained Residual Encoder U-Net. The segmentation performance was subsequently evaluated by employing the Dice similarity coefficient and Jaccard's Index, also known as the Intersection Over Union (IoU) metric. A 5-fold cross-validation approach was used in the evaluation, to assess the consistency of the network. This approach allowed us to validate the trained models against diverse subsets of the data, mitigating potential biases associated with any single fold. Out of the 5 folds, the network was validated on two folds with the highest results being reported.

\begin{figure}
\includegraphics[width=\textwidth]{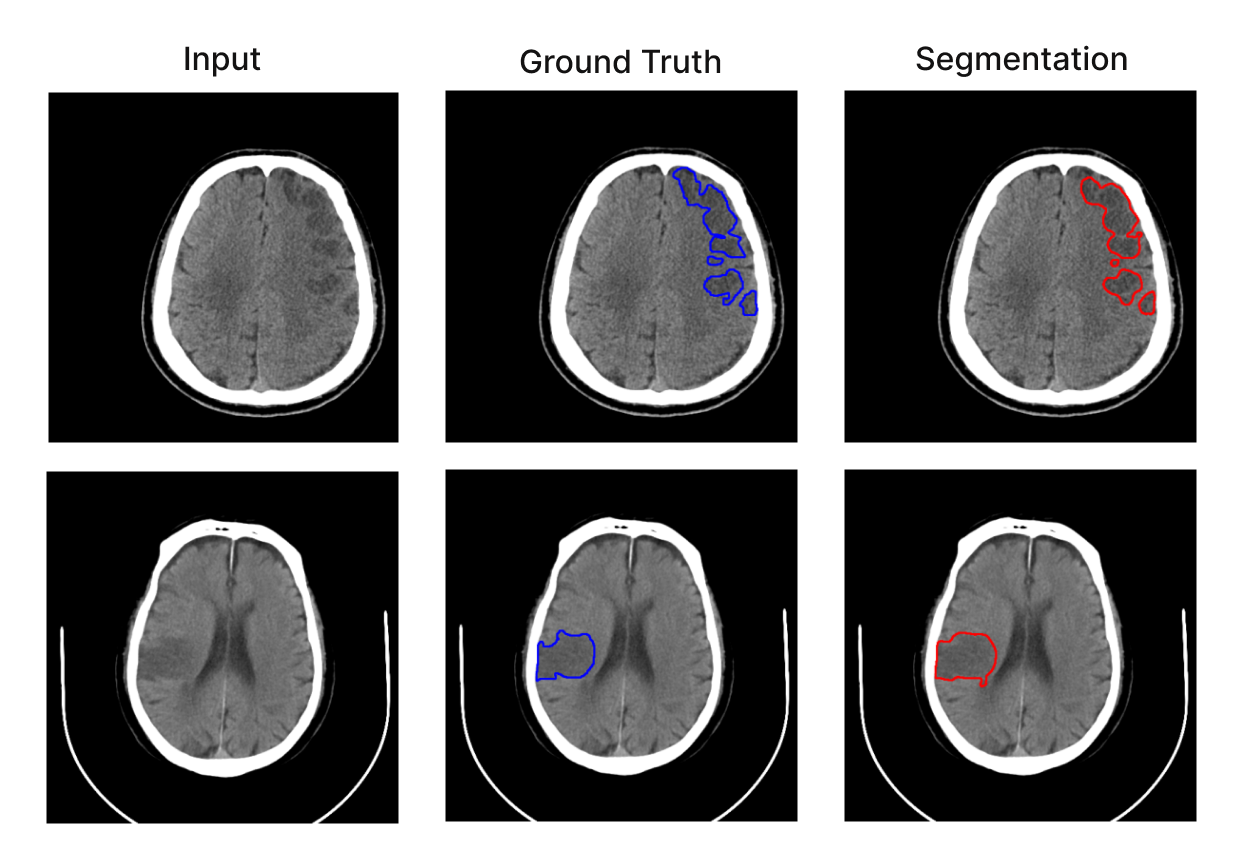}
\caption{Comparing the ground truth segmentation results in blue, to the network's segmentation results in red} \label{fig:gt-vs-seg}
\end{figure}

The mean Dice score achieved was 0.596, with a Jaccard's index of 0.510 as seen in table \ref{tab:performance_metrics}. Compared to the Symmetry-Enhanced Attention Network's \cite{Liang_2021} reported dice score of 0.578 on the same dataset, these results can be viewed as promising in enabling the segmentation of acute infarctions from NCCT scans. Scores ranged from as low as 0.00, to as high as 0.923, which was attributable to the existence of several outliers in the dataset. The outliers included volume regions located outside the typical areas associated with stroke, extending to regions such as the top and base of the skull. Without the outliers, the resulting dice score is as high as 0.752, with a Jaccard's index of 0.643.

\begin{table}[H]
\caption{Comparison of Segmentation Performance Metrics}\label{tab:performance_metrics}
\centering
\setlength{\tabcolsep}{10pt}
\begin{tabular}{lcc}
\hline
\textbf{Metric} & \textbf{Value} & \textbf{Description} \\
\hline
Mean Dice Score & 0.596 & Overall mean score \\
Jaccard's Index & 0.510 & Overall mean index \\
Dice Score Range & 0.00 - 0.983 & Extremes due to outliers \\
Adjusted Mean Dice Score & 0.752 & Excluding outliers \\
Adjusted Jaccard's Index & 0.643 & Excluding outliers \\
Comparative Dice Score & 0.578 & SEAN \cite{Liang_2021} \\
\hline
\end{tabular}
\end{table}

\noindent The relatively low Dice scores observed in the segmentation task can largely be attributed to the inherent challenges in tissue differentiation within NCCT scans \cite{Difficult-CT}, compounded by the presence of several outliers in the dataset. Despite these challenges, NCCT remains an essential diagnostic tool for distinguishing between hemorrhagic and ischemic strokes, with the latter presenting more complexities in accurate detection. The results are promising and underscore the need for further data collection to enhance the effectiveness of segmentation algorithms in infarct segmentation using NCCT scans. The additional data could potentially refine the training process, leading to improvements in model performance and, consequently, more accurate clinical assessments for better patient outcomes.

\section{Conclusion}
In this paper, we employ nnU-Net's residual encoder U-Net architecture in the automatic segmentation of acute stroke infarctions in NCCT scans. Our investigations revealed that while NCCT poses significant challenges in ischemic lesion identification due to low tissue differentiation, the method demonstrated potential in segmenting the lesions with a reasonable degree of accuracy. Notably, the Dice scores, although relatively low, showed promising avenues for future improvements.

Accurate delineation of the infarction region is pivotal for clinical decision-making. It enables clinicians to evaluate the extent and severity of the stroke, providing essential insights into the potential impact on brain function. Detailed infarct delineation supports targeted treatment strategies, such as the administration of thrombolytics for ischemic strokes or surgical interventions in more severe cases. While challenges persist, the improvements in automated segmentation of stroke lesions in NCCT scans hold considerable promise for improving stroke diagnosis and treatment, ultimately enhancing patient care and outcomes.

\bibliographystyle{splncs04}
\bibliography{references}

\end{document}